\documentclass[vecphys]{svmult}
\RequirePackage{graphicx}
\RequirePackage{makeidx}
\RequirePackage{multicol}
\usepackage[bottom]{footmisc}

\begin{document}

\title{Coherent optical pulse dynamics in nanocomposite plasmonic Bragg
gratings}
\titlerunning{Coherent optical pulse dynamics in nanocomposite...}
\author{Ildar\,R.\,Gabitov\inst{1,2} \and
    Alexander\,O.\,Korotkevich\inst{2} \and
    Andrei\,I.\,Maimistov\inst{3} \and
    Joseph\,B.\,McMahon\inst{4}}
\authorrunning{Ildar\,R.\,Gabitov et al.}
\institute{Department of Mathematics, University of Arizona\\
617 North Santa Rita Avenue, Tucson, AZ 85721, USA\\
\texttt{gabitov@math.arizona.edu} \and
L.D. Landau Institute for Theoretical Physics, Russian Academy of Sciences\\2 Kosygin Street,
Moscow, 119334, Russian Federation\\
\texttt{kao@itp.ac.ru} \and
Department of Solid State Physics, Moscow Engineering Physics
Institute\\Moscow, 115409, Russian Federation\\
\texttt{maimistov@pico.miphi.ru} \and
Program in Applied Mathematics, University of Arizona\\617 North Santa Rita Avenue,
P.O. Box 210089, Tucson, AZ 85721-0089, USA\\
\texttt{jmcmahon@math.arizona.edu}}

\maketitle

\begin{abstract}The propagation of solitary waves in a Bragg grating
formed by an array of thin nanostructured dielectric films is
considered. A system of equations of Maxwell-Duffing type and describing
forward- and backward-propagating waves in such a grating, is
derived. Exact solitary wave solutions are found, analyzed,  and
compared with the results of direct numerical simulations.
\end{abstract}

\section{Introduction}

The last decade has been a period of rapid progress in the field of photonic
crystals~\cite{1}-\cite{10}. In particular, the one-dimensional case of a
\textit{resonant Bragg grating} \cite{1}-\cite{5} or a \textit{resonantly
absorbing Bragg reflector} (RABR)~\cite{6}-\cite{8} has been studied extensively.
In the simplest case a resonant Bragg grating consists of a linear
homogeneous dielectric medium containing an array of thin films with
resonant atoms or molecules. The thickness of each film is much less than the wavelength of the
electromagnetic wave propagating through such a structure. The interaction
of ultra-short pulses and films embedded with two-level atoms has been
studied by Mantsyzov et al.~\cite{1}-\cite{5} in the framework of the
two-wave reduced Maxwell-Bloch model and by Kozhekin et al.~\cite{6}-\cite{8}.
This work demonstrated the existence of the $2\pi$-pulse of self-induced
transparency in such structures~\cite{1,4,6}. It was also
found~\cite{8} that bright as well as dark solitons can exist in the
prohibited spectral gap, and that bright solitons can have arbitrary pulse
area.

If the density of two-level atoms is very high, then the near-dipole-dipole
interaction is noticeable and should be accounted for in the mathematical
model. The effect of dipole-dipole interaction on the existence of gap
solitons in a resonant Bragg grating was studied in~\cite{10}; details can
be found in~\cite{9}. Recent numerical simulations have yielded unusual
solutions known as \emph{zoomerons}.~\cite{CD80} The optical zoomeron was
discovered and investigated recently \cite{M05} in the context of the
resonant Bragg grating. A zoomeron is a localized pulse similar to an
optical soliton, except that its velocity oscillates about some mean value.
These works also contain careful construction of the underlying mathematical
model, which is derived from first principles.

Recent advances in nanofabrication have allowed the creation of
nanocomposite materials, which have the ability to sustain nonlinear
plasmonic oscillations. These materials have metallic nanoparticles embedded
in them~\cite{R97}-\cite{HRF86}. In this paper we consider a
dielectric material into which thin films containing metallic nanoparticles
have been inserted. These thin films are spaced periodically along the
length of the dielectric so that the Bragg prohibited spectral gap is
centered at the plasmonic resonance frequency of the nanoparticles. We
derive governing equations for slowly-varying envelopes of two
counter-propagating electromagnetic waves and of the plasmonic
oscillation-induced medium polarization. We find that this system of
equations has the form of the two-wave Maxwell-Duffing model. We find exact
solutions of this system and demonstrate that, in contrast to conventional
$2\pi$-pulses, they have nonlinear phase. We show that the stability of these
solutions is sensitive to perturbation of this phase. We also study the
collisions of these pulses and find that the outcomes of such collisions
are highly dependent on relative phase.

\section{Basic Equations}

We consider a grating formed by an array of thin films which are
embedded in a linear dielectric medium. In our derivation of the
governing equations we follow~\cite{1}-\cite{8}, wherein Bragg
resonance arises if the distance between successive films is
$a = \left(\lambda/2\right)m$, $m = 1, 2, 3, \ldots$~. To obtain
governing equations we apply the transfer-operator approach, presented
below.

\subsection{ Transfer-operator approach}

Let us consider the ultra-short optical pulse propagation along the
$X$-direction of the periodic array of thin films, which are placed
at points $\ldots, x_{n-1}, x_{n}, x_{n+1},\ldots $ (Fig.1). The
medium between films has dielectric permittivity
$\varepsilon$. Hereafter, for the sake of definiteness, we consider a TE-wave
whose electric field component is parallel to the layers.
All results can be generalized easily for the case of TM-polarized waves.

It is suitable to represent the electric and magnetic strengths
$\vec{E}, \vec{H}$, and the polarization
of the two-level atoms ensemble $\vec{P}$ in the form of Fourier integrals
\begin{eqnarray*}
\vec{E}(x,z,t) & = & (2\pi )^{-2}\int\limits_{-\infty}^{\infty}\exp [-i\omega t+i\beta z]\vec{E}(x,\beta ,\omega )dt~dz,\\
\vec{H}(x,z,t) & = & (2\pi )^{-2}\int\limits_{-\infty }^{\infty}\exp [-i\omega t+i\beta z]\vec{H}(x,\beta ,\omega )dt~dz,\\
\vec{P}(x_{n},z,t) & = & (2\pi )^{-2}\int\limits_{-\infty}^{\infty }\exp [-i\omega t+i\beta z]\vec{P}(x_{n},\beta,\omega )dt~dz.
\end{eqnarray*}

Outside the films the Fourier components of the vectors
$\vec{E}(x,\beta ,\omega)$ and
$\vec{H}(x,\beta ,\omega )$\
are defined by the Maxwell equations. At points $x_{n}$ these values
are defined from continuity conditions. Thus, the TE-wave
propagation can be described by the following system.
\begin{equation}
\frac{d^{2}E}{dx^{2}}+\left( k^{2}\varepsilon -\beta ^{2}\right)E =0, \label{eq1a}
\end{equation}
\begin{eqnarray*}
H_{x} = -(\beta /k)E,\quad H_{z} = -(i/k)dE/dx,\quad E_{y} = E,
\end{eqnarray*}
with boundary conditions \cite{R11,R12}
\begin{equation}
E(x_{n}-0)=E(x_{n}+0),\quad H_{z}(x_{n}+0)-H_{z}(x_{n}-0)=4i\pi
kP_{y}(x_{n},\beta ,\omega ),  \label{eq1b}
\end{equation}
where $k = \omega /c$. The solutions of equation (\ref{eq1a}) in the intervals
$x_{n} < x < x_{n+1}$ can be written as
\begin{eqnarray*}
E(x,\beta ,\omega ) & = & A_{n}(\beta ,\omega )\exp
[iq(x-x_{n})]+B_{n}(\beta ,\omega )\exp [-iq(x-x_{n})],\\
H_{z}(x,\beta ,\omega ) & = & qk^{-1}\left\{ A_{n}(\beta ,\omega )\exp
[iq(x-x_{n})]-B_{n}(\beta ,\omega )\exp [-iq(x-x_{n})]\right\},
\end{eqnarray*}
where $q=\sqrt{k^{2}\varepsilon -\beta ^{2}}$ . Hence, the
amplitudes $A_{n}$ and $B_{n}$ completely determine the
electromagnetic field in a RABR. Let us
consider the point $x_{n}$. The electric field at
$x=x_{n}-\delta $ ($\delta <<a$) is defined by amplitudes
$A_{n}^{(L)}$\ and $B_{n}^{(L)}$, and the field at
$x=x_{n}+\delta $ is defined by $A_{n}^{(R)}$\ and $B_{n}^{(R)}$.
Continuity conditions (\ref{eq1a}) result in the following relations
among these amplitudes
\begin{eqnarray*}
A_{n}^{(R)}\ + B_{n}^{(R)} & = & A_{n}^{(L)}\ +B_{n}^{(L)},\\
A_{n}^{(R)}\ -B_{n}^{(R)}=A_{n}^{(L)}\ -B_{n}^{(L)}+4\pi ik^{2}q^{-1}P_{S,n},
\end{eqnarray*}
where $P_{S,n}=P_{S}(A_{n}^{(R)}\ +B_{n}^{(R)})$ is the surface
polarization of a thin film at point $x_{n}$, which is induced by the
electrical field inside the film. Thus we find
\begin{equation}
A_{n}^{(R)}\  =  A_{n}^{(L)}+2\pi ik^{2}q^{-1}P_{S,n},\quad
B_{n}^{(R)}\  =  B_{n}^{(L)}-2\pi ik^{2}q^{-1}P_{S,n}.\label{eq2}
\end{equation}
Taking into account the strength of the electric field outside the films, we write
\begin{equation}
A_{n+1}^{(L)}\ = A_{n}^{(R)}\exp (iqa),\quad B_{n+1}^{(L)}\
= B_{n}^{(R)}\exp (-iqa).  \label{eq3}
\end{equation}

If the vectors $\psi _{n}^{(L)}=(A_{n}^{(L)},B_{n}^{(L)})$ and
$\psi_{n}^{(R)}=(A_{n}^{(R)},B_{n}^{(R)})$
are introduced, then the relations (\ref{eq2}) can be represented as
\[
\psi _{n}^{(R)} = \widehat{U}_{n}\psi _{n}^{(L)},
\]
where $\widehat{U}_{n}$\ is the \emph{transfer operator} of vector
$\psi_{n}^{(L)}$ through the film located at point $x_{n}$. In the general case
$\widehat{U}_{n}$\ is a nonlinear operator.
The relations (\ref{eq3}) are represented in the vectorial form
\[
\psi _{n+1}^{(L)}=\widehat{V}_{n}\psi _{n}^{(R)},
\]
where the linear operator $\widehat{V}_{n}$ transfers the vector
$\psi_{n}^{(R)}$ between adjacent thin films and is represented by the diagonal matrix
\[
\widehat{V}_{n}=\left(
\begin{array}{cc}
\exp (iqa) & 0 \\
0 & \exp (-iqa)
\end{array}
\right).
\]
In this manner we define the nonlinear transfer-operator of the vector
$\psi_{n}^{(L)}$ through an elementary cell of RABR:
\begin{equation}
\psi _{n+1}^{(L)}=\widehat{V}_{n}\widehat{U}_{n}\psi _{n}^{(L)}
= \widehat{T}_{n}\psi _{n}^{(L)}.  \label{eq4}
\end{equation}
The transfer-operator approach is frequently used in models of one-dimensional
photonic crystals of linear media, e.g. distributed feedback structures \cite{R13}.

In (\ref{eq4}) the upper index can be omitted, and the equation can be
rewritten as the following recurrence relations
\begin{eqnarray}
A_{n+1} & = & A_{n}\exp (iqa)+2\pi ik^{2}q^{-1}P_{S,n}\exp (iqa),\label{eq5a}\\
B_{n+1} & = & B_{n}\exp (-iqa)-2\pi ik^{2}q^{-1}P_{S,n}\exp (-iqa)\label{eq5b}
\end{eqnarray}
These recurrence relations are exact, as no approximations (e.g.
slowly-varying envelope of electromagnetic pulses approximation, the
long-wave approximation) have been employed. Furthermore, the
surface polarization of a thin film could be calculated via
different suitable models. Here we follow the works by B. Mantsyzov
at al. \cite{1}-\cite{5}, and A. Kozhekin, G. Kurizki, et al.
\cite{6}-\cite{9}, where the two-level atom model has been used.

\subsection{Linear response approximation }

To demonstrate that the RABR is a true gap medium, it is suitable to
obtain the electromagnetic wave spectrum through a linear response
approximation. In the general case we can use the following expression for polarization:
\begin{equation}
P_{S,n}=\chi (\omega )(A_{n}^{(R)}\ +B_{n}^{(R)}).  \label{eq6}
\end{equation}
Substitution of this formula in (\ref{eq5a}),(\ref{eq5b}) yields
\begin{eqnarray}
A_{n+1} & = & (1+i\rho )A_{n}\exp (iqa)+i\rho B_{n}\exp (iqa),\label{eq7a}\\
B_{n+1} & = & (1-i\rho )B_{n}\exp (-iqa)-i\rho A_{n}\exp (-iqa).\label{eq7b}
\end{eqnarray}
Here $\rho =\rho (\omega )=2\pi k^{2}q^{-1}\chi (\omega )=2\pi
\omega c^{-1}\varepsilon ^{-1/2}\chi (\omega )$. We employ an ansatz
in which the wave is a collective motion of the electrical field in the
grating.  Hence
\begin{equation}
A_{n} = A\exp (ikna),\qquad B_{n}=B\exp (ikna).  \label{eq8}
\end{equation}
Refs. (\ref{eq7a}),(\ref{eq7b}) show that the wave amplitudes $A$ and $B$ satisfy the following linear system of equations:
\begin{eqnarray}
A\exp (iKa) & = & (1+i\rho )A\exp (iqa)+i\rho B\exp (iqa),  \label{eq9a}\\
B\exp (iKa) & = & (1-i\rho )B\exp (-iqa)-i\rho A\exp (-iqa).  \label{eq9b}
\end{eqnarray}
A nontrivial solution of this system exists if and only if the determinant is equal to zero, i.e.
\begin{equation}
\det \left(
\begin{array}{cc}
(1+i\rho )\exp (iqa)-\exp (iKa) & i\rho \exp (iqa) \\
-i\rho \exp (-iqa) & (1-i\rho )\exp (-iqa)-\exp (iKa)
\end{array}
\right) = 0.  \label{eq10}
\end{equation}

If we define
\begin{eqnarray*}
Z & = & \exp (iKa),\\
G & = & (1+i\rho )\exp (iqa) = (\cos qa-\rho \sin qa)+i(\rho \cos qa+\sin qa),
\end{eqnarray*}
then equation (\ref{eq10}) can be rewritten as the following equation in $Z$:
\[
Z^{2}-(G+G^{\ast })Z+1=0.
\]
This equation has solutions
\[
Z_{\pm }=\mathrm{Re~} G\pm i\sqrt{1-(\mathrm{Re~} G)^{2}}.
\]

If $\mathrm{Re~} G\leq 1$, then\
$\mathrm{Re~} G\pm i\sqrt{1-(\mathrm{Re~} G)^{2}}=\cos Ka+i\sin Ka$.
Hence the wave numbers $K_{\pm }$ are real-valued and satisfy
the transcendental equation
\begin{equation}
\cos Ka = \cos qa - \rho \sin qa.  \label{eq11}
\end{equation}
\begin{figure}[hbt]
\centering
\includegraphics[width=12cm]{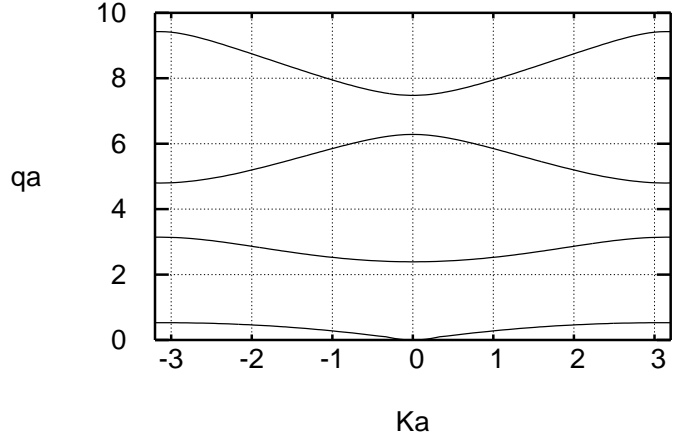}
\caption{An example of band structure, if inclusions are linear
oscillators. } \label{bandstructure}
\end{figure}
%This dispersion relation defines the dependence of wave number
%$k_{\pm }$ on the frequency of a harmonic wave propagating in a linear RABR.

If $\mathrm{Re~} G > 1$, then the roots of equation (\ref{eq10}) are
real. In this case, the wave numbers $K_{\pm }$ are pure imaginary.
The condition $\mathrm{Re~} G > 1$ defines the frequencies of the
forbidden zone. The waves with these frequencies cannot propagate in
the grating. The boundaries of this forbidden zone are defined by
\begin{equation}
\cos qa - \rho \sin qa = 1.  \label{eq12}
\end{equation}

The model of the resonant system containing the thin films defines
the explicit form of the function $\rho =\rho (\omega )$.  The form
of $\rho (\omega)$ determines the dispersion relation (\ref{eq11}).
It should be noted that this dispersion relation ensures a series of
gaps in the electromagnetic wave spectrum.
Figure~(\ref{bandstructure}) represents an example of such a band
structure when the inclusions are linear oscillators.

\subsection{  Long-wave and weak nonlinearity approximations }

By using the long-wave approximation we can transform the exact equations
(\ref{eq5a}), (\ref{eq5b}) into differential equations. To do so
we introduce the field variables
\[
A(x) = \sum\limits_{n}A_{n}\delta (x-x_{n}),\quad
B(x) = \sum\limits_{n}B_{n}\delta (x-x_{n}),\quad P(x) = \sum\limits_{n}P_{S,n}\delta (x-x_{n}),
\]
Using the integral representation for a Dirac delta-function,
\[
\delta (x)=(2\pi )^{-1}\int\limits_{-\infty }^{\infty }\exp (ikx)dk,
\]
we produce the following expression
\begin{eqnarray*}
A(x) & = &(2\pi )^{-1}\sum_{n}A_{n}\int\limits_{-\infty }^{\infty}\exp[ik(x-x_{n})]dk \\
& = &(2\pi )^{-1}\int\limits_{-\infty }^{\infty }\exp(ikx)\sum_{n}A_{n}\exp (-ikx_{n})dk.
\end{eqnarray*}
It follows that the Fourier transform of $A(x)$ is then
\[
A(k)=\sum_{n}A_{n}\exp (-ikx_{n})=\sum_{n}A_{n}\exp (-ikan),
\]
and the Fourier transforms of $B(x)$ and $P(x)$ are
\[
B(k)=\sum_{n}B_{n}\exp (-ikan),\quad P(k)=\sum_{n}P_{S,n}\exp
(-ikan),
\]
\noindent respectively.

The form of these spatial Fourier components ensures periodicity in $k$. For example,
\begin{eqnarray}
A(k) & = &\sum_{n}A_{n}\exp (-ikan)\exp (\pm 2\pi in)  \nonumber \\
& = &\sum_{n}A_{n}\exp (-ikan\pm 2\pi in)  \label{eq14} \\
& = &\sum_{n}A_{n}\exp [-ian(k\pm 2\pi /a)] = A(k\pm 2\pi /a). \nonumber
\end{eqnarray}

>From recurrence equations (\ref{eq5a}),(\ref{eq5b}) we have
\begin{eqnarray*}
A(k)\exp (ika) & = & A(k)\exp (iqa)+i\kappa P(k)\exp (iqa),\\
B(k)\exp (ika) & = & B(k)\exp (-iqa)-i\kappa P(k)\exp (-iqa),
\end{eqnarray*}
or
\begin{eqnarray}
\left[ \exp \{ia(k-q)\}-1\right] A(k) & = & i\kappa P(k),  \label{eq15a}\\
\left[ \exp \{ia(k+q)\}-1\right] B(k) & = & -i\kappa P(k).  \label{eq15b}
\end{eqnarray}

The linear approximation $P(k)=\chi (\omega)[A(k)+B(k)]$\
for polarization again leads us to the linear dispersion law (\ref{eq10}).
The system of equations (\ref{eq15a}),(\ref{eq15b}) is equivalent to the discrete equations
(\ref{eq5a}),(\ref{eq5b}). At this point we have made no approximations
other than the assumption on the thin films' width.

If the thin film array were absent, then the dispersion relation would be
\[
\cos ka = \cos qa.
\]
In this case the wave with amplitude $A(k)$ has $k=q$, indicating propagation to the right,
while the wave with amplitude $B(k)$ has $k=-q$,
indicating propagation in the opposite direction.
Let us suppose that the polarization of the thin film array
produces little change in wave vectors, i.e.,
for the right-propagating wave the wave vector is
$k = q + \delta k$, and for the opposite wave the
wave vector is $k = -q + \delta k$.  If we set the
value of $q$ near one of the Bragg resonances,
say $q=2\pi /a+\delta q$, where $\delta q\ll 2\pi /a$,
then equations (\ref{eq5a}), (\ref{eq5b})
take the form
\begin{eqnarray*}
\left[ \exp (ia\delta k)-1\right] A(q+\delta k) & = & i\kappa P(q+\delta k),\\
\left[ \exp (ia\delta k)-1\right] B(-q+\delta k) & = & -i\kappa P(-q+\delta k).
\end{eqnarray*}
By virtue of the periodicity conditions (\ref{eq14}) these equations can be re-written as
\begin{eqnarray*}
\left[ \exp (ia\delta k)-1\right] A(\delta q+\delta k) & = & i\kappa P(\delta q+\delta k),\\
\left[ \exp (ia\delta k)-1\right] B(-\delta q+\delta k) & = & -i\kappa
P(-\delta q+\delta k).
\end{eqnarray*}
After the change of variables $\delta k=\delta \tilde{k}\pm \delta q$ we have
\begin{eqnarray}
\left[ \exp \{ia(\delta \tilde{k}-\delta q)\}-1\right] A(\delta \tilde{k})
& = & i\kappa P(\delta \tilde{k}),  \label{eq16a}\\
\left[ \exp \{ia(\delta \tilde{k}+\delta q)\}-1\right] B(\delta \tilde{k})
& = & -i\kappa P(\delta \tilde{k}).  \label{eq16b}
\end{eqnarray}
The long-wave approximation means that nonzero values of the spatial
Fourier amplitudes are located near zero value of argument. Suppose that
$a\delta \tilde{k}$ is small enough that
$e^{ia(\delta \tilde{k}+\delta q)} \approx 1 + ia(\delta \tilde{k}+\delta q)$.
Then  we replace equations (\ref{eq16a}), (\ref{eq16b}) with the approximate equations
\begin{equation}
ia(\delta \tilde{k}-\delta q)A(\delta \tilde{k})=i\kappa P(\delta
\tilde{k}), \label{eq17a}
\end{equation}
\begin{equation}
ia(\delta \tilde{k}+\delta q)B(\delta \tilde{k})=-i\kappa P(\delta \tilde{k}).  \label{eq17b}
\end{equation}
If now we return to the spatial variable, equations (\ref{eq17a}), (\ref{eq17b}) lead us to the equations of coupled-wave theory
\begin{eqnarray}
\frac{\partial A}{\partial x} & = & i\delta qA(x)+i\kappa a^{-1}P(x),  \label{eq18a}\\
\frac{\partial B}{\partial x} & = & -i\delta qB(x)-i\kappa a^{-1}P(x).  \label{eq18b}
\end{eqnarray}

In these equations the fields $A(x), B(x), P(x)$ and the parameters $\delta q$,
$\kappa $ are functions of frequency $\omega $. To obtain the
final system of equations in the spatial and time variables, we perform an
inverse Fourier transformation.  We now assume that the envelopes of the
electromagnetic waves vary slowly in time \cite{R14}. This
approximation simplifies the system of coupled wave
equations under consideration.

\subsection{  Slowly-varying envelope approximation }

In the slowly-varying envelope approximation, we assume that the approximated fields
are inverse Fourier transforms of narrow wave packets, i.e. they are quasiharmonic waves \cite{R14}.
For example, the quasiharmonic wave form of the electric field is
\[
E(x,t) = \mathcal{E}(x,t)\exp [-i\omega _{0}t],
\]
where $\omega _{0}$ is carrier wave frequency. The electric field
$E(x,t)$ and the Fourier components of the envelope $\mathcal{E}(x,t)$ of the pulse
are related by
\begin{eqnarray*}
E(x,t) & = &(2\pi )^{-1}\int\limits_{-\infty }^{\infty }\widetilde{E}(x,\omega)\exp(-i\omega t)d\omega\\
& = & (2\pi )^{-1}\int\limits_{-\infty }^{\infty }\widetilde{\mathcal{E}}(x,\omega)\exp[-i(\omega +\omega _{0})t]d\omega\\
& = & (2\pi )^{-1}\int\limits_{-\infty }^{\infty }\widetilde{\mathcal{E}}(x,\omega-\omega _{0})\exp (-i\omega t)d\omega,
\end{eqnarray*}
where the function $\widetilde{E}(x,\omega )$ is nonzero if
$\omega \in(\omega _{0}-\Delta \omega ,\omega _{0}+\Delta \omega )$,
with $\Delta \omega \ll \omega _{0}$.\ Hence,
$\widetilde{E}(x,\omega +\omega _{0}) = \widetilde{\mathcal{E}}(x,\omega )$.
Thus, if we have some relation for $\widetilde{E}(x,\omega )$, then the analogous relation for
$\widetilde{\mathcal{E}}(x,\omega )$ can be found by shifting $\omega \mapsto \omega _{0}+\omega $
in all functions of $\omega $.

Let
\[
A(x,t) = \mathcal{A}(x,t)\exp (-i\omega _{0}t),\quad B(x,t)=\mathcal{B}(x,t)\exp (-i\omega _{0}t),
\quad P(x,t) = \mathcal{P}(x,t)\exp (-i\omega _{0}t).
\]
>From Eqs. (\ref{eq18a}), (\ref{eq18b}) it follows that the Fourier components
$\mathcal{A}(x,\omega),\mathcal{B}(x,\omega )$,
and $\mathcal{P}(x,\omega )$ satisfy
\begin{eqnarray}
\frac{\partial \mathcal{A}}{\partial x}(x,\omega ) & = & i\delta q(\omega _{0}+\omega )
\mathcal{A}(x,\omega )+i\kappa (\omega _{0}+\omega )a^{-1}\mathcal{P}
(x,\omega ),  \label{eq19a}\\
\frac{\partial \mathcal{B}}{\partial x}(x,\omega ) & = & -i\delta q(\omega _{0}+\omega )
\mathcal{B}(x,\omega )-i\kappa (\omega _{0}+\omega )a^{-1}\mathcal{P}
(x,\omega )  \label{eq19b}
\end{eqnarray}
Since $\mathcal{A}$, $\mathcal{B}$ and $\mathcal{P}$
\ are nonzero for $\omega \ll \omega _{0}$,
one can use the expansions:
\begin{equation}
\delta q(\omega _{0}+\omega )\approx q_{0}-2\pi /a+q_{1}\omega
+q_{2}\omega ^{2}/2,\quad \kappa (\omega _{0}+\omega )a^{-1}\approx
K_{0}, \label{eq20}
\end{equation}
where $q_{n}=d^{n}q/d\omega ^{n}$ at $\omega =\omega _{0}$. In
particular, $q_{1}^{-1} = v_{g}$ is the group velocity, and $q_{2}$
takes into account the group-velocity dispersion.

Considering expansions (\ref{eq20}), we have the following description
of the evolution of slowly-varying envelopes.
\begin{eqnarray}
i\left( \frac{\partial }{\partial x}+\frac{1}{v_{g}}\frac{\partial }{\partial t}\right) \mathcal{A}-\frac{q_{2}}{2}\frac{\partial ^{2}}{\partial
t^{2}}\mathcal{A}+\Delta q_{0}\mathcal{A} & = & -K_{0}\mathcal{P},
\label{eq21a}\\
i\left( \frac{\partial }{\partial x}-\frac{1}{v_{g}}\frac{\partial }{\partial t}\right) \mathcal{B}+\frac{q_{2}}{2}\frac{\partial ^{2}}{\partial
t^{2}}\mathcal{B}-\Delta q_{0}\mathcal{B} & = & +K_{0}\mathcal{P},
\label{eq21b}
\end{eqnarray}
where $\Delta q_{0}=q_{0}-2\pi /a$. The next step requires a choice of
the model for the thin films' medium. Possibilities include anharmonic
oscillators, two- or three-level atoms, excitons of molecular chains,
nano-particles, quantum dots, and more. First we consider the two-level atom model.

\subsection{ Example:  thin films containing two-level atoms }

Here we employ the approach developed above to derive an already
known system of equations~\cite{1}.  The model assumes that each
thin film contains two-level atoms. The state of a two-level atom is
described by a density matrix $\hat{\rho}$. The matrix element $\rho
_{12}$ describes the transition between the ground state $\left\vert
2\right\rangle $ and excited state $\left\vert 1\right\rangle$.
$\rho _{22}$ and $\rho _{11}$ represent the populations of theses
states. Evolution of the two-level atom is governed by the Bloch
equations \cite{R15}.
\begin{eqnarray}
i\hbar \frac{\partial }{\partial t}\rho _{12} & = &
\hbar \Delta \omega\rho _{12}-d_{12}(\rho _{22}-\rho _{11})A_{in},
\label{eq22a}\\
i\hbar \frac{\partial }{\partial t}(\rho _{22}-\rho _{11}) & = &
2\left(d_{12}\rho _{21}A_{in}-d_{21}\rho _{12}A_{in}^{\ast }\right) .
\label{eq22b}
\end{eqnarray}
In these equations $A_{in}$ is the electric field interacting with
a two-level atom. In the problem under consideration,
$A_{in}=\mathcal{A}+\mathcal{B}$, and
\[
K_{0}\mathcal{P}=\frac{2\pi \omega _{0}n_{at}d_{12}}{cn(\omega _{0})}
\left\langle \rho _{12}\right\rangle .
\]
Here the cornerstone brackets denote summation over all atoms
within a frequency detuning of $\Delta \omega $ from the center of the
inhomogeneity broadening line, $n(\omega _{0})$ is the refractive
index of the medium containing the array of thin films, and $n_{at}$ is
the effective density of the resonant atoms in the films. $n_{at}$ is
defined by $n_{at}=N_{at}(\ell_{f}/a)$, where $N_{at}$ is the bulk density of atoms,
$\ell_{f}$ is the film width, and $a$ is the lattice spacing.

We suppose the group-velocity dispersion is of no importance.
The resulting equations are the two-wave reduced Maxwell-Bloch equations.
We introduce the normalized variables
\[
e_{1}=t_{0}d_{12}\mathcal{A}/\hbar ,\quad
e_{2}=t_{0}d_{12}\mathcal{B}/\hbar ,\quad
x=\zeta v_{g}t_{0} ,\quad
\tau =t/t_{0}.
\]
The normalized two-wave reduced Maxwell-Bloch equations take the following form:
\begin{eqnarray}
i\left( \frac{\partial }{\partial \zeta }+\frac{\partial }{\partial \tau }\right) e_{1}+\delta e_{1}
& = & -\gamma \left\langle \rho_{12}\right\rangle , \label{eq23a}\\
i\left( \frac{\partial }{\partial \zeta }-\frac{\partial }{\partial \tau }
\right) e_{2}-\delta e_{2}
& = & +\gamma \left\langle \rho_{12}\right\rangle , \label{eq23b}\\
i\frac{\partial }{\partial \tau }\rho _{12} & = &
\Delta \rho_{12}-ne_{in}, \label{eq23c}\\
\frac{\partial }{\partial \tau }n & = &
-4~\mathrm{Im~} (\rho_{12}e_{in}^{\ast }), \label{eq23d}
\end{eqnarray}
where $\gamma =t_{0}v_{g}/L_{a}$, $\delta =t_{0}v_{g}\Delta q_{0}$, $
L_{a}=(cn(\omega _{0})\hbar )/(2\pi \omega_{0}t_{0}n_{at}|d_{12}|^{2})$\
is the resonant absorption length, and
$\Delta =\Delta \omega t_{0}$\ is the normalized frequency detuning.

We define $n = \rho _{22}-\rho _{11}$, $e_{in} = e_{1} + e_{2}$ and
introduce yet another change of variables:
\[
e_{in}=e_{1}+e_{2}=f_{s}\exp (i\delta \tau ),\quad
e_{1}-e_{2}=f_{a}\exp (i\delta \tau ),\quad \rho _{12}=r\exp
(i\delta \tau ).
\]
The system of equations (\ref{eq23a}) - (\ref{eq23d}) can be rewritten as
\begin{eqnarray}
\frac{\partial f_{s}}{\partial \zeta } + \frac{\partial f_{a}}{\partial \tau } & = &
0,  \label{eq25a}\\
\frac{\partial f_{a}}{\partial \zeta } + \frac{\partial f_{s}}{\partial \tau } & = &
2i\gamma \left\langle r\right\rangle ,  \label{eq25b}\\
i\frac{\partial }{\partial \tau }r & = &
(\Delta +\delta )r-nf_{s}, \label{eq25c}\\
\frac{\partial }{\partial \tau }n & = &
-4~\mathrm{Im~} (rf_{s}^{\ast }). \label{eq25d}
\end{eqnarray}
>From (\ref{eq25a}) it follows that
\[
\frac{\partial f_{a}}{\partial \zeta } = -\frac{\partial f_{s}}{\partial \tau },
\]
which allows to rewrite (\ref{eq25a})-(\ref{eq25d}) in the form
\begin{eqnarray}
\frac{\partial ^{2}f_{s}}{\partial \zeta ^{2}}-\frac{\partial ^{2}f_{s}}{\partial \tau ^{2}} & = &
-2i\gamma \left\langle \frac{\partial r}{\partial \tau }\right\rangle ,  \label{eq26a}\\
i\frac{\partial }{\partial \tau }r & = &
(\Delta +\delta )r-nf_{s}, \label{q26b}\\
\frac{\partial }{\partial \tau }n & = &
-4~\mathrm{Im~} (rf_{s}^{\ast }). \label{eq26c}
\end{eqnarray}

If we assume that the inhomogeneous broadening is absent, i.e. if
the hypothesis of a sharp atomic resonant transition is true, then
$\delta +\Delta = 0$, and this system reduces to the Sine-Gordon
equation \cite{1}. Reference \cite{4} presents the steady-state
solution of (\ref{eq26a})-(\ref{eq26c}) with inhomogeneous
broadening taken into account.

\subsection{Thin films containing metallic nanoparticles}

It was shown above that counter-propagating electric field waves
$\mathcal{A}$ and $\mathcal{B}$ in
the slowly-varying envelope approximation satisfy the
following system of equations:
\begin{eqnarray}
i\left( \frac{\partial }{\partial x}+\frac{1}{v_{g}}\frac{\partial }{
\partial t}\right) \mathcal{A}-\frac{q_{2}}{2}\frac{\partial ^{2}}{\partial
t^{2}}\mathcal{A}+\Delta q_{0}\mathcal{A} &=&-\frac{2\pi \omega
_{0}}{c\sqrt{
\varepsilon }}\langle\mathcal{P}\rangle,  \label{21.1} \\
i\left( \frac{\partial }{\partial x}-\frac{1}{v_{g}}\frac{\partial }{
\partial t}\right) \mathcal{B}+\frac{q_{2}}{2}\frac{\partial ^{2}}{\partial
t^{2}}\mathcal{B}-\Delta q_{0}\mathcal{B} &=&+\frac{2\pi \omega
_{0}}{c\sqrt{ \varepsilon }}\langle\mathcal{P}\rangle,  \label{21.2}
\end{eqnarray}
where $\Delta q_{0}=q_{0}-2\pi /a$ is the mismatch between the
carrier wavenumber and the Bragg resonant wavenumber. To describe
the evolution of material polarization in the slowly-varying
amplitude approximation, we must model the thin films' response to
an external light field. Previous work has considered various
mechanisms as sources of the dielectric properties of metamaterials.
In the simplest case, dielectric properties can be attributed to
plasmonic oscillations, which are modeled by Lorentz oscillators.
Magnetic properties can be described by the equations of a system of
LC-circuits~\cite{PHRS99,SSMS02,KKKES04,MS02,WWM05}. The simplest
generalizations of this model include anharmonicity of plasmonic
oscillations~\cite{R16,LGMS} or the addition of a nonlinear
capacitor into each LC-circuit~\cite{ZSK03}. In this paper we
consider an array of non-magnetic thin films containing metallic
nanoparticles, which have cubic nonlinear response to external
fields~\cite{R97,DBNS04,R16}.

The macroscopic polarization $P$ is governed by the equation
\[
\frac{\partial ^{2}P}{\partial t^{2}}+\omega _{d}^{2}P+\Gamma
_{a}\frac{\partial P}{\partial t}+\kappa P^{3}=\frac{\omega _{p}^{2}}{4\pi }E,
\]
where $\omega _{p}$ is plasma frequency and $\omega _{d}$ is
dimension quantization frequency for nanoparticles. Losses of the
plasmonic oscillations are taken into account by the parameter
$\Gamma_{a}$. It is assumed that the duration  of the
electromagnetic pulse is small enough that dissipation effects can
be neglected. If the anharmonic parameter $\kappa$ is equal to zero,
then we have the famous Lorentz model for describing electromagnetic
wave propagation and refraction in
metamaterials~\cite{Zi26,Zi27,PHRS99,SSMS02,KKKES04,MS02,WWM05}.

Starting from the slowly-varying envelope approximation,
standard manipulation leads to
\begin{equation}
i\frac{\partial \mathcal{P}}{\partial t}+(\omega _{d}-\omega_{0})\mathcal{P}
+ \frac{3\kappa }{2\omega_{0}}|\mathcal{P}|^{2}\mathcal{P} =
-\frac{\omega _{p}^{2}}{8\pi\omega _{0}}\mathcal{E}_{int}(x,t).  \label{anhar3}
\end{equation}
Terms varying rapidly in time, which are proportional to $\exp (\pm 3i\omega
_{0}t)$, are neglected. In this equation $\mathcal{E}_{int}$ is the
electric field interacting with metallic nanoparticles. In the problem under
consideration we have $\mathcal{E}_{int}=\mathcal{A}+\mathcal{B}$ .

Due to the limitations of nanofabrication, the sizes and shapes of
nanoparticles are not uniform. In practice, deviation from a perfectly
spherical shape has a much larger impact on a nanoparticle's resonance
frequency than does variation in diameter. This causes a broadening of the
resonance line. The broadened spectrum is characterized by a probability
density function $g(\Delta \omega )$ of deviations $\Delta \omega $ from
some mean value $\omega_{res}$. When computing the total polarization, all
resonance frequencies must be taken into account.

The contributions of the various resonance frequencies are weighted according to the
probability density function $g(\Delta \omega )$; the weighted
average is denoted by $\langle\mathcal{P}\rangle$ in
equations~(\ref{21.1}),(\ref{21.2}).
In what follows, $n(\omega _{0})$ denotes the refractive index of
the medium containing the array of thin films, and $n_{np}$
is the effective density of the resonant nanoparticles in films.
As in the model of films containing two-level atoms,
the effective density is equal to
$n_{np} = N_{np}(\ell_{f}/a)$, where $N_{np}$ is the bulk density
of nanoparticles, $\ell_{f}$ is the width of a film, and $a$ is the lattice spacing.

We study a medium-light interaction in which resonance is the
dominant phenomenon. As such, the length of the sample is smaller
than the characteristic dispersion length. In this case the temporal
second derivative terms in equations~(\ref{21.1},\ref{21.2}) can be
omitted. The resulting equations are the two-wave
Maxwell-Duffing equations. They can be rewritten in dimensionless
form using the following rescaling:
\begin{eqnarray*}
e_{1} & = &\mathcal{A}/A_{0}, \\
e_{2} & = &\mathcal{B}/A_{0}, \\
p & = & (4\pi \omega _{0}/[\sqrt{\varepsilon }\omega _{p}A_{0}])\mathcal{P}, \\
\zeta  & = &(\omega _{p}/2c)x, \\
\tau  & = &t/t_{0}.
\end{eqnarray*}
Here $t_{0}=2\sqrt{\varepsilon }/\omega _{p}$, while $A_{0}$  is
a characteristic amplitude of counter-propagating fields. In dimensionless
form, the two-wave Maxwell-Duffing equations read
\begin{eqnarray}
i\left( \frac{\partial }{\partial \zeta }+\frac{\partial }{\partial
\tau }
\right) e_{1}+\delta e_{1} &=&-\langle p\rangle ,  \nonumber \\
i\left( \frac{\partial }{\partial \zeta }-\frac{\partial }{\partial
\tau }
\right) e_{2}-\delta e_{2} &=&+\langle p\rangle ,\label{SVEPEquations} \\
i\frac{\partial p}{\partial \tau }+\Delta p+\mu |p|^{2}p
&=&-(e_{1}+e_{2}),\nonumber
\end{eqnarray}
where $\mu =(3\kappa \sqrt{\varepsilon }/\omega
_{0}\omega_{p})(\sqrt{ \varepsilon }\omega _{p}/4\pi
\omega_{0})^{2}A_{0}^{2}$ is a dimensionless coefficient of
anharmonicity, $\delta =2\Delta q_{0}(c/\omega _{p})$ is the
dimensionless mismatch coefficient, $\Delta
=2\sqrt{\varepsilon}(\omega _{d}-\omega _{0})/\omega _{p}$ is the
dimensionless detuning of a nanoparticle's resonance frequency from
the field's carrier frequency.

In a coordinate system rotating with angular frequency $\delta$,
\[
e_{1}=f_{1}e^{i\delta \tau },\;\indent e_{2}=f_{2}e^{i\delta \tau },\;p=qe^{i\delta\tau },
\]
equations~(\ref{SVEPEquations}) become
\begin{eqnarray}
i\left( \frac{\partial }{\partial \zeta }+\frac{\partial }{\partial\tau }
\right) f_{1} &=&-\langle q\rangle ,  \nonumber \\
i\left( \frac{\partial }{\partial \zeta }-\frac{\partial }{\partial\tau }
\right) f_{2} &=&+\langle q\rangle , \label{RotatingFrame}\\
i\frac{\partial q}{\partial \tau }+(\Delta -\delta )q+\mu |q|^{2}q
& = &-(f_{1}+f_{2}).\nonumber
\end{eqnarray}
Further simplification of the system~(\ref{RotatingFrame}) can be
achieved by introducing new variables
\[
f_{s}=-(f_{1}+f_{2}),\indent f_{a} = f_{1}-f_{2},
\]
which allow decoupling of one equation from the system of three
equations. In these new variables the polarization $q$ is coupled with
only one field variable. Simple transformations give

\begin{eqnarray}
\frac{\partial ^{2}f_{a}}{\partial \zeta ^{2}}-
\frac{\partial ^{2}f_{a}}{\partial \tau ^{2}} & = & 2i
\frac{\partial }{\partial \zeta }\langle q\rangle ,\label{26.3}\\
\frac{\partial ^{2}f_{s}}{\partial \zeta ^{2}}-
\frac{\partial ^{2}f_{s}}{\partial \tau ^{2}} & = & 2i
\frac{\partial }{\partial \tau }\langle q\rangle,\label{26.3.1}\\
i\frac{\partial q}{\partial \tau }+(\Delta -\delta )q +
\mu |q|^{2}q & = & f_{s}.\label{26.3.2}
\end{eqnarray}
As one can see, we have a coupled system of equations for $f_{s}$ and $q$.

\section{Solitary Wave Solutions}

We consider localized solitary wave solutions of~(\ref{26.3}) in the
limit of narrow spectral line $\Delta \omega_{g}/\Delta \omega_{s}
\ll 1$,where $\Delta \omega_{s}$ and  $\Delta \omega_{g}$ are
spectral widths of a signal and spectral line $g(\Delta\omega)$. In this
case the spectral line can be represented as Dirac $\delta$-function:
$g(\Delta\omega)=\delta(\Delta\omega)$. Equations~(\ref{26.3}),(\ref{26.3.2}) can then be
re-written as follows:
\begin{eqnarray}
\frac{\partial ^{2}f_{s}}{\partial \zeta ^{2}}-
\frac{\partial ^{2}f_{s}}{\partial \tau ^{2}} & = & 2i\frac{\partial q}{\partial \tau }\\
i\frac{\partial q}{\partial \tau }+(\Delta -\delta )q +
\mu |q|^{2}q & = & f_{s}
\end{eqnarray}
Scaling analysis  of this system shows that solitary wave solutions
can be represented as
\begin{eqnarray}
f_{s}&=& f_0 F_{\Omega}\left(\eta \right) = \frac{1}{\sqrt{\mu}}\left( \frac{2 v^2}{1-v^2}\right
)^{3/4}F_{\Omega}\left(\eta \right),\nonumber\\
q&=& q_0 Q_{\Omega}\left(\eta \right) = \frac{1}{\sqrt{\mu}}\left( \frac{2 v^2}{1-v^2}\right
)^{1/4}Q_{\Omega}\left(\eta \right),\label{Scaling}\\
\eta & = & (\zeta - v\tau)\sqrt{\frac{2}{1-v^2}}.\nonumber
\end{eqnarray}
Here $v$ is velocity of the solitary wave, $\eta$ is a
scale-invariant parameter in a coordinate system moving with the
solitary wave, and functions $F_{\Omega}$, $Q_{\Omega}$ satisfy the
following system of equations:
\begin{eqnarray}
F'' & = & -iQ',\\
-iQ' + \Omega Q + |Q|^2 Q & = & F.
\end{eqnarray}
The only dimensionless parameter which remains in the system is
\begin{eqnarray}
\Omega = (\Delta - \delta)\sqrt{\frac{1 - v^2}{2 v^2}},
\end{eqnarray}
which characterizes the deviation of carrier frequency from the plasmonic frequency
$\omega_p$ and the Bragg resonance frequency $\omega_{Br}$.

The first equation implies that $F' = -iQ + \textrm{constant}$.   We
seek a solitary-wave solution, so we assume that $F$, $Q$, and their
derivatives decay to zero as $|\eta|\to\infty$.  Hence the constant
is zero, and
\begin{eqnarray}
iF' &=& Q \nonumber\\
-iQ' + \Omega Q + |Q|^2 Q  &=&  F.\label{ODE}
\end{eqnarray}

The system of ordinary differential equations~(\ref{ODE}) has integral
of motion
\[
|Q|^2 - |F|^2 = \mathrm{constant}.
\]
For solutions decaying as $|\eta|\to\infty$, the constant  is equal
to zero, and
\[
|Q|^2 = |F|^2.
\]
This allows the following parametrization of solutions:
\[
F(\eta) = R(\eta) e^{i\phi(\eta)},\indent Q (\eta)= R(\eta)
e^{i\psi(\eta)},
\]
where $R$, $\phi$, and $\psi$ are real-valued functions satisfying
\begin{eqnarray}
R' & = & -R\sin(\phi - \psi),\nonumber\\
\phi' & = & -\cos(\phi - \psi),\label{AmplitudePhase}\\
\psi' + \Omega + R^2 & = & \cos(\phi - \psi).\nonumber
\end{eqnarray}
If we set $\Phi = \phi - \psi$, then we have
\begin{eqnarray}
\Phi' - \Omega - R^2 & = & -2\cos\Phi,\nonumber\\
R' & = & -R\sin\Phi.\label{AmplitudePhaseCompact}
\end{eqnarray}
Taking into account second equation
of~(\ref{AmplitudePhaseCompact}),  the first equation can be
re-written as
\[
R\frac{d}{dR}\left(\cos\Phi\right) - \Omega - R^2 = -2\cos\Phi.
\]
If we set $y = \cos\Phi$, then we have
\begin{equation}
R\frac{dy}{dR} + 2y = R^2 + \Omega.\label{EquationForY}
\end{equation}
The solutions of equation~(\ref{EquationForY}) have the form
\[
y = \frac{1}{4}R^2 + \frac{\Omega}{2} + cR^{-2},
\]
where $c$ is free.  Since $y = \cos\Phi$, the right-hand
side must remain between -1 and 1.  As we expect $R\to 0$ as
$|\eta|\to\infty$, $c$ must be zero.  We have the conservation law
\begin{equation}
\cos\Phi = \frac{1}{4}R^2 +
\frac{\Omega}{2}.\label{SecondConsevationLaw}
\end{equation}
Substituting~(\ref{SecondConsevationLaw}) into the second  equation
of~(\ref{AmplitudePhaseCompact}) and subsequent integration gives
the following expression for $R$:
\[
R^2 = \frac{2(4 - \Omega^2)}{\Omega + 2\cosh\left\{ \sqrt{4 -
\Omega^2}(\eta - \eta_0) +
\frac{1}{2}\ln\left(\frac{16}{4-\Omega^2}\right)\right\}}.
\]
The right-hand side is positive real-valued for all
$\eta$ if and only if $-2 < \Omega < 2$.  $\sqrt{4-\Omega^2}$
appears so often in what follows that we set $\beta =
\sqrt{4-\Omega^2}$.  $R^2$ then has the form
\begin{equation}
\label{r_solution}
R^2 = \frac{2\beta^2}{\Omega + 2\cosh\left\{\beta(\eta - \eta')\right\}},
\end{equation}
where we have combined arbitrary constant $\eta_0$ and  the
logarithm into a single constant ($\eta'$) in the argument of the
hyperbolic cosine. Using the conservation
law~(\ref{SecondConsevationLaw}) we obtain an expression for $\Phi$:
\begin{equation}
\label{Phi_expression}
\Phi = 2\arctan\left(\frac{2-\Omega}{\beta}\tanh\left\{
\frac{1}{2}\beta(\eta - \eta')\right\}\right).
\end{equation}
Now we integrate $\phi' = -\cos\Phi$ and find
\begin{eqnarray*}
\phi & = & -\frac{\Omega}{2}(\eta - \eta') - \arctan\left(
\frac{2-\Omega}{\beta}\tanh\left\{\frac{1}{2}\beta(\eta -
\eta')\right\}\right).
\end{eqnarray*}
Finally, we determine $\psi$:
\begin{eqnarray*}
\psi & = & \phi - \Phi \nonumber\\
& = & -\frac{\Omega}{2}(\eta - \eta') -
3\arctan\left(\frac{2-\Omega}{\beta}\tanh\left\{\frac{1}{2}\beta(\eta - \eta')\right\}\right).
\end{eqnarray*}
This pulse exists only if    value of the parameter $\Omega$ are
inside the interval $-2<\Omega<2$. The maximal value of the
amplitude of this solitary solution is
\[
A = \sqrt{2(2-\Omega)}.
\]
The phases $\phi$ and $\psi$ are nonlinear. Their behavior is
asymptotically linear as $\eta \rightarrow \pm\infty$. If
$\Omega = 0$, then the limiting values of the phases satisfy
\begin{eqnarray*}
|\phi (\infty)-\phi(-\infty)| & = & \pi/2,\label{fieldphase}\\
|\psi(\infty)-\psi(-\infty)| & = & 3\pi/2.
\end{eqnarray*}

\section{Energy Partition}

The total energy of the solitary wave is distributed among
co/contr-propagating fields and medium polarization.  Here we study
energy partition between all these components.  Using equations
(\ref{26.3}), (\ref{26.3.1}) and conditions as
$|\eta|\rightarrow\infty$, one can show that
\begin{equation}
f_a(\eta) = -\frac{1}{v} f_s(\eta) = -\frac{f_0}{v}F(\eta).
\end{equation}
We are interested in the energies of the dimensionless fields $f_1$,
$f_2$, and polarization $q$.
\begin{eqnarray*}
f_1 & = & \frac{1}{2}(f_a - f_s) = -\frac{f_0}{2}\left(\frac{1}{v} + 1\right)F\\
f_2 & = & -\frac{1}{2}(f_s + f_a) = \frac{f_0}{2}\left(\frac{1}{v} - 1\right)F\\
q & = & q_0 Q
\end{eqnarray*}
To find energies of forward- and backward-propagating  waves we need
to calculate an energy of a solitary wave:
\begin{eqnarray}
E_R ~=~ \int\limits_{-\infty}^{+\infty}|F|^2(\eta)d\eta ~=~ \int\limits_{-\infty}^{\infty}|Q|^2(\eta)d\eta
~=~\nonumber\\
\int\limits_{-\infty}^{+\infty}R^2(\eta)d\eta ~=~
8\arctan\sqrt{\frac{2-\Omega}{2+\Omega}}.
\end{eqnarray}
Finally we have the energies
\begin{eqnarray}
E_{f_1} &=& \frac{f_0^2}{4}\left(\frac{1}{v} + 1\right)^2 E_R,\\
E_{f_2} &=& \frac{f_0^2}{4}\left(\frac{1}{v} - 1\right)^2 E_R,\\
E_q &=& q_0^2 E_R.
\end{eqnarray}
Ratios of energies in different fields as well as polarization have
the following form
\begin{eqnarray}
\frac{E_{f_1}}{E_{f_2}} &=& \left(\frac{1+v}{1-v}\right)^2,\\
\frac{E_{f_1}}{E_q} &=& \frac{1}{2}\frac{(1+v)}{(1-v)},\\
\frac{E_{f_2}}{E_q} &=& \frac{1}{2}\frac{(1-v)}{(1+v)}.
\end{eqnarray}
Therefore, energy partitioning is determined by only one parameter
$v$ which is dimensionless combination of main system parameters.
\section{Numerical simulation}

The shape and phase of the incident pulse are controllable in  real
experimental situation. To model pulse dynamics in the Bragg grating
it is natural to consider asymptotic mixed initial-boundary  value
problem for equations~(\ref{RotatingFrame}).  We define initial
conditione as
\begin{equation}
q(\zeta,\tau)\rightarrow 0,~~~f_1(\zeta,\tau)\rightarrow
0,~~~f_2(\zeta,\tau)\rightarrow 0,~~~\tau \rightarrow -\infty
\end{equation}
with no incident field at the left edge of the sample and with
incident field at the left edge defined as follows:
\begin{eqnarray}
&f_1&(-10,\tau) = w\exp(i\theta),\nonumber\\
&w& = 3.5\exp\left[-\frac{1}{2}\left(\frac{\tau -3.0}{1.5}\right)^2\right], \\
&\theta& = \arctan\left(\tanh\left[1.5\left(\tau -
3.0\right)\right]\right).\nonumber
\end{eqnarray}
In our case the spatial simulation  domain  was chosen as
$[-10,40]$. Parameters $\Delta -\delta$ and $\mu$ were  chosen
\begin{equation}
\Delta -\delta = 0,\;\; \mu = 1.
\end{equation}

As one can see we gave the initial pulse the same configuration of
phase in topological sense as in solitary wave solution. This point
is important, because otherwise phase difference cannot relax to the
symmetry of the stationary wave which is revealed
in~(\ref{Phi_expression}). As a result without right ``topological
charge'' solution will be unstable.

For the second field boundary condition was as follows
\begin{equation}
f_2 (40,\tau) = 0.
\end{equation}

In the first experiment we injected a pulse relatively close to
solitary wave solution. The results are shown in
Figs.~\ref{f1.1.map}-\ref{q.1.map}.
\begin{figure}[hbt]
\centering
\includegraphics[width=4.0in]{figures/bin.E1.colour.exact.Gauss.3.5.eps2}
\caption{Propagation of pulse. The first experiment. Mapping of the
$|e_1 (\zeta,\tau)|$ surface.} \label{f1.1.map}
\end{figure}
\begin{figure}[hbt]
\centering
\includegraphics[width=4.0in]{figures/bin.E2.colour.exact.Gauss.3.5.eps2}
\caption{Propagation of pulse. The first experiment. Mapping of the
$|e_2 (\zeta,\tau)|$ surface.} \label{f2.1.map}
\end{figure}
\begin{figure}[hbt]
\centering
\includegraphics[width=4.0in]{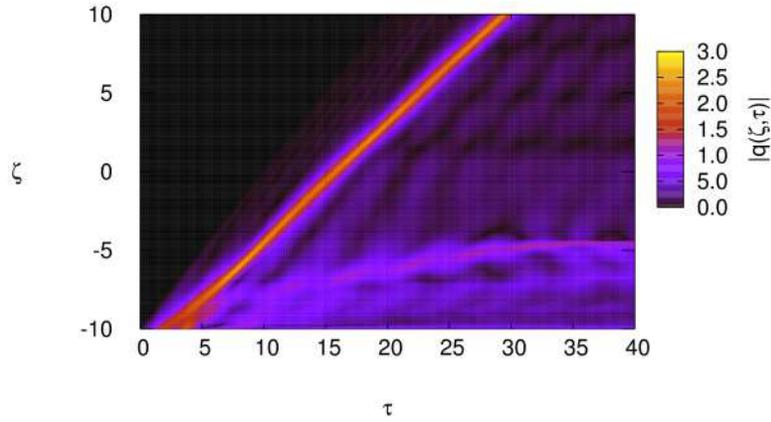}
\caption{Propagation of pulse. The first experiment. Mapping of the
$|q(\zeta,\tau)|$ surface.} \label{q.1.map}
\end{figure}
The amplitude phase difference slightly differed as well as pulse
shape which was Gaussian. On the first stage ($t \le 7$) of
evolution we observed fast excess energy damping in radiation of
quasi-linear waves in both directions and relaxation to solution
roughly close to stationary one. Then we had some stage of pulse
shape refinement ($ 7 < t < 30$) with consequent propagation of the
solution very close to (\ref{r_solution}). One can compare refined
pulse shape with stationary solution in Fig.~\ref{q.1.contour}.
\begin{figure}[hbt]
\centering
\includegraphics[width=10.0cm]{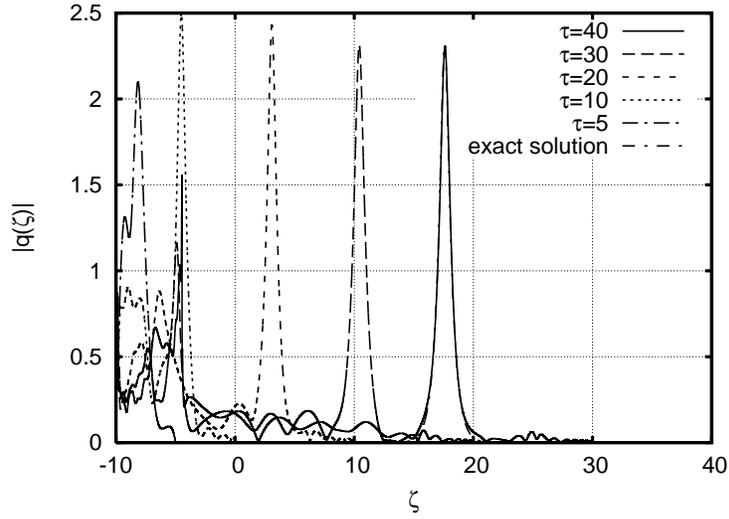}
\caption{The first experiment. Absolute value of polarization at
different moments of time. Stationary pulse shape almost coincides
with exact solution.} \label{q.1.contour}
\end{figure}

During second experiment we used pulse of the lower amplitude
\begin{eqnarray}
&f_1&(-10,\tau) = w\exp(i\theta),\nonumber\\
&w& = 2.0\exp\left[-\frac{1}{2}\left(\frac{\tau -3.0}{1.5}\right)^2\right], \\
&\theta& = \arctan\left(\tanh\left[1.5\left(\tau - 3.0\right)\right]\right).
\end{eqnarray}
Results are represented in Figs.~\ref{f1.2.map}-\ref{q.2.map}.
\begin{figure}[hbt]
\centering
\includegraphics[width=4.0in]{figures/bin.E1.colour.exact.Gauss.2.0.eps2}
\caption{Propagation of pulse. The second experiment. Mapping of the
$|e_1 (\zeta,\tau)|$ surface.} \label{f1.2.map}
\end{figure}
\begin{figure}[hbt]
\centering
\includegraphics[width=4.0in]{figures/bin.E2.colour.exact.Gauss.2.0.eps2}
\caption{Propagation of pulse. The second experiment. Mapping of the
$|e_2 (\zeta,\tau)|$ surface.} \label{f2.2.map}
\end{figure}
\begin{figure}[hbt]
\centering
\includegraphics[width=4.0in]{figures/bin.Q.colour.exact.Gauss.2.0.eps2}
\caption{Propagation of pulse. The second experiment. Mapping of the
$|p(\zeta,\tau)|$ surface.} \label{q.2.map}
\end{figure}
Clearly observable oscillations of pulse magnitude can be associated
with modes localized on pulse. This very instructive result could be
a fruitful ground for further investigation and experiments.

\section{Conclusion}
We derived equations describing optical pulse evolution in Bragg
gratings with thin films containing active dopants. In particular we
investigated case of thin films containing metallic nanoparticles.
We showed that corresponding system have one parameter. This
parameter contains information on Bragg and plasmonic frequencies
and general pulse characteristics.  These equations have   solitary
wave solutions describing bound state of two contraindicating waves
with medium polarization.

\section*{Acknowledgments}

We would like to thank  B.\,I.~Mantsyzov,  A.\,A.~Zabolotskii, J-G.~Caputo,
M.\,G.~Stepanov and R.~Indik for enlightening discussions. AIM
and KAO are grateful to the Laboratoire de Math\'{e}matiques, INSA
de Rouen and the University of Arizona for hospitality and support.
This work was partially supported by NSF (grant DMS-0509589),
ARO-MURI award 50342-PH-MUR  and State of Arizona (Proposition 301),
RFBR grants 06-02-16406 and 06-01-00665-a, INTAS grant 00-292, the
Programme ``Nonlinear dynamics and solitons'' from the RAS Presidium
and ``Leading Scientific Schools of Russia'' grant. KAO was
supported by Russian President grant for young scientists
MK-1055.2005.2.


\begin{thebibliography}{99}
\bibitem{1} B.\,I. Mantsyzov and R.\,N. Kuzmin, Sov. Phys. JETP \textbf{64},
37-44 (1986).

\bibitem{2} B.\,I. Mantsyzov and D.\,O. Gamzaev, Optics and spectroscopy
\textbf{63}, 1, 200-202 (1987).

\bibitem{3} T.\,I. Lakoba and B.I.Mantsyzov, Bull. Russian Acad. Sci., Phys.
\textbf{56}, 8, 1205-1208 (1992).

\bibitem{4} B.\,I. Mantsyzov, Phys.Rev. \textbf{A51}, 6, 4939-4943 (1995).

\bibitem{5} B.\,I. Mantsyzov and E.\,A. Silnikov, J. Opt. Soc. Amer. B,
\textbf{19}, 2203-2207 (2002).

\bibitem{6} A. Kozhekin and G. Kurizki, Phys. Rev. Lett. \textbf{74}, 25,
5020-5023 (1995).

\bibitem{7} A. Kozhekin, G. Kurizki, and B. Malomed, Phys. Rev. Lett.
\textbf{81}, 17, 3647-3650 (1998).

\bibitem{8} T. Opatrny, B.\,A. Malomed, and G. Kurizki, Phys. Rev. E \textbf{
60}, 5, 6137-6149 (1999).

\bibitem{9} G. Kurizki, A.\,E. Kozhekin, T. Opatrny, and B.\,A. Malomed,
Progr. Optics \textbf{42}, 93-146 (E. Wolf, editor: North Holland,
Amsterdam, 2001)

\bibitem{10} J. Cheng, J. Zhou, Phys.Rev. E \textbf{66}, 036606 (2002).


\bibitem{R11} V.I. Rupasov, V.I. Yudson, \textit{Quantum electronics} (in
Russia) \textbf{9}, 2179 (1982).

\bibitem{R12} V.I. Rupasov, V.I. Yudson, \textit{Zh.E.T.Ph.}(in Rusia)
\textbf{93}, 494 (1987).

\bibitem{R13} P. Yeh, \textit{Optical Waves in Layered Media} (Wiley, New
York, 1988).

\bibitem{R14} A.I. Maimistov, A.M. Basharov, \textit{Nonlinear Optical Waves
} (Dordrecht, Boston, London: Kluwer Academic Publishers, 1999).

\bibitem{R15} L. Allen, J.H. Eberly, \textit{Optical resonance and two-level
atoms}, Wiley-Interscience, New York, 1975.

\bibitem{R16} I.R. Gabitov, R.A. Indik, N.M. Litchinitser, A.I. Maimistov,
V.M. Shalaev, J.E. Soneson, \textit{J. Opt. Soc. Am}. \textbf{B.23},
535-542 (2006)


\bibitem{MaPo} A.\thinspace I. Maimistov and V.\thinspace V. Polikarpov,
Quantum Electronics (in Russian) \textbf{36}, 835 (2006).
\bibitem{R18} A.A. Zabolotskii, private communication.


\bibitem{CD80} F. Calogero, A. Degasperis, in Solitons, R. K. Bullough, P.
J. Caudray, eds. (Springer-Verlag, Berlin, 1980).

\bibitem{M05} B.\,I. Mantsyzov,   "Optical zoomeron as a result
of beatings of the internal   modes of a Bragg soliton", JETP
letters, \textbf{82}, 5, 253-258 (2005).

\bibitem{R97} S. G. Rautian, ``Nonlinear saturation spectroscopy of the
degenerate electron gas in spherical metallic particles'',\ JETP
\textbf{85}, 451--461 (1997).

\bibitem{DBNS04} V. P. Drachev, A. K. Buin, H. Nakotte, and V. M. Shalaev,
``Size dependent $\chi^3$ for conduction electrons in Ag
nanoparticles'',\ Nano Lett. \textbf{4}, 1535--1539 (2004).

\bibitem{HRF86} F. Hache, D. Ricard, and C. Flytzanis, ``Optical
nonlinearities of small metal particles: surface-mediated resonance
and quantum size effects'',\ J. Opt. Soc. Am. B \textbf{3},
1647--1655 (1986).

\bibitem{SS94} C. M. De Sterke, J. E. Sipe, Prog. Opt. \textbf{33}, 203
(1994).

\bibitem{K99} R. Kashyap, Fiber Bragg Gratings (San Diego: Academic Press,
1999).

\bibitem{KA05} G. Agrawal, Y. S. Kivshar, Optical Solitons: From Fibers to
Photonic Crystals (Amsterdam: Academic Press. 2003).

\bibitem{PHRS99} J. B. Pendry, A. J. Holden, D. J. Robbins, W. J. Stewart,
IEEE Trans. Microwave Theory Techniques \textbf{47}, 2075-2084
(1999).

\bibitem{SSMS02} D. R. Smith, S. Schultz, P. Markos, C. M. Soukoulis, Phys.
Rev. B \textbf{65}, 195104 (2002).

\bibitem{KKKES04} N. Katsarakis, T. Koschny, M. Kafesaki, E. N. Economou, C.
M. Soukoulis, Appl. Phys. Lett. \textbf{84}, 2943-2945 (2004).

\bibitem{MS02} P. Markos, C. M. Soukoulis, Phys. Rev. E \textbf{65}, 036622
(2002).

\bibitem{WWM05} J. F. Woodley, M. S. Wheeler, M. Mojahedi, Phys. Rev. E
\textbf{71}, 066605 (2005).

\bibitem{LGMS} N. M. Lichinitser, I. R. Gabitov, A. I. Maimistov, V. M.
Shalaev, \textquotedblleft Effect of an optical negative index thin
film on optical bistability\textquotedblright , arXiv.
physics/0607177.

\bibitem{ZSK03} A. A. Zharov, I. V. Shadrivov, Yu. S. Kivshar, Phys. Rev.
Lett. \textbf{91}, 037401 (2003).

\bibitem{Ma06} A.\thinspace I. Maimistov , ``On Coherent Optical Pulse
Propagation in One-Dimensional Bragg Grating'', \ arXiv.
nlin.PS/0607012

\bibitem{Zi26} R.W. Ziolkowskii, E. Heyman. Phys. Rev. B \textbf{64}  056625
(2001).

\bibitem{Zi27} R.W. Ziolkowski, Optics Express 11, {2116}7, 662 - 681
(2003).

\end{thebibliography}
\end{document}